\documentclass[aps,prl,reprint,showpacs,showkeys,superscriptaddress]{revtex4-1}
\usepackage{graphicx}
\usepackage{bm}
\usepackage{color}
\usepackage{amsmath}
\usepackage{amsfonts}
\usepackage{amssymb}
\usepackage{natbib}
\usepackage{latexsym}

\begin{document}

{\raggedleft {\it Accepted for publication in Physical Review Letters}}
\\

\title{Relativistic Magnetic Reconnection in Kerr Spacetime}
\author{Felipe A. Asenjo}
\email{felipe.asenjo@uai.cl}
\affiliation{Facultad de Ingenier\'{\i}a y Ciencias, Universidad Adolfo Ib\'a\~nez, Santiago 7941169, Chile.}
\author{Luca Comisso}
\email{lcomisso@princeton.edu}
\affiliation{Department of Astrophysical Sciences and Princeton Plasma Physics Laboratory, Princeton University, Princeton, NJ 08544, USA.}


\begin{abstract}
The magnetic reconnection process is analyzed for relativistic magnetohydrodynamical plasmas around rotating black holes. A simple generalization of the Sweet-Parker model is used as a first approximation to the problem. The reconnection rate, as well as other important properties of the reconnection layer, have been calculated taking into account the effect of spacetime curvature. Azimuthal and radial current sheet configurations in the equatorial plane of the black hole have been studied, and the case of small black hole rotation rate has been analyzed. For the azimuthal configuration, it is found that the  black hole  rotation decreases the reconnection rate. On the other hand, in the radial configuration, it is the gravitational force created by the black hole mass that  decreases the reconnection rate. These results establish a fundamental interaction between gravity and magnetic reconnection in astrophysical contexts.
\end{abstract}


\maketitle

{\it Introduction.-} The rapid conversion of magnetic energy into plasma particle energy through the process of magnetic reconnection is of paramount importance in many astrophysical processes \cite{Kulsrud_2005}. Magnetospheric substorms, coronal mass ejections, stellar and gamma-ray flares are just a few examples of pheneomena in which magnetic reconnection is thought to play a crucial role.

In recent years, significant work has been undertaken to better understand magnetic reconnection in magnetically dominated environments, where relativistic effects become significant \cite{Hoshino_2012}. This has led to the generalization of the classical Sweet-Parker and Petschek reconnection models to the special relativistic regime \cite{lyu,luca1}. Furthermore, many numerical campaigns have been devoted to the investigation of the reconnection rate \cite{Zenitani2009,Kowal2009,Zenitani2010,Takahashi_2011,Bessho2012,Takamoto2013,Liu2015,Takamotoin} and the particle acceleration \cite{Zenitani2007,kowalparticle1,kowalparticle2,Cerutti2012,Sironi2014,Guo2014,Guo2015,Werner2016} in this regime.
Relativistic reconnection has been found to be a very efficient mechanism of magnetic energy conversion and particle acceleration, making it a primary candidate to explain nonthermal emissions from pulsar wind nebulae, gamma-ray bursts, and active galactic nuclei.

While special relativistic effects on the reconnection process are starting to become clear, the effects related to the spacetime curvature are far less explored and a detailed understanding is lacking. General relativistic magnetohydrodynamic simulations have repeatedly shown the formation of reconnection layers in proximity of blacks holes \cite{Koide2006,Penna2010,LyutMcK_2011,McKinney2012}, where spacetime curvature effects can be important. However, the difficulties related to the spatial and temporal resolution of typical reconnection processes have been such to prevent their thorough study.

A step forward in the comprehension of magnetic reconnection in curved spacetime could be taken by studying simple generalizations of known theoretical models. It is the purpose of this Letter to develop such theoretical study considering the contribution of the gravitational field of a rotating black hole on the magnetic reconnection process. Ultimately, our goal is to determine how the reconnection rate, and other properties of the reconnection layer, are modified by spacetime curvature effects.

{\it Governing equations.-}
We consider a plasma governed by the equations of General Relativistic Magnetohydrodynamics (GRMHD) \cite{lich,Anile_1989}, which are composed by the continuity equation  $\nabla_\nu \left( \rho U^\nu \right) = 0$, the energy-momentum equation 
\begin{equation}\label{eqmom}
\nabla_\nu\left(\mathfrak{h} U^\mu U^\nu\right)=-\nabla^\mu p+J^\nu {F^{\mu}}_\nu\, , 
\end{equation}
and the resistive Ohm's law
\begin{equation}\label{eqOhm}
U^\nu {F^\mu }_\nu  = \eta (J^\mu - \rho_e' U^\mu) \, .
\end{equation}
Here, $\nabla_\nu$ denotes the covariant derivative associated to the connections of the curved spacetime, $U^\mu$ and $J^\mu$ are the four-velocity and four-current density, respectively, while $F^{\mu\nu}$ is the electromagnetic field tensor. Furthermore, $\rho$, $\mathfrak{h}$ and $p$ are the mass density, enthalpy density, and pressure of the plasma.
Finally, $\eta$ is the electrical resistivity and $\rho_e'= - U^\nu J_\nu$ is the charge density observed by the local center-of-mass frame of the plasma. The description of the dynamics is then completed when the fluid equations are complemented by Maxwell's equations $\nabla_\nu F^{\mu\nu} = J^\mu$ and $\nabla_\nu F^{*\mu\nu}=0$, where $F^{*\mu\nu}$ is the dual of the electromagnetic field tensor.

A very effective representation of these equations can be obtained by writing them in the $3+1$ formalism \cite{TM_95}. In this case, the curvature effects  are displayed explicity in a set of vectorial GRMHD equations. In this formulation, the line element can be written as 
\begin{equation} 
d{s^2} = g_{\mu \nu} d{x^\mu}d{x^\nu} =  - {\alpha ^2}d{t^2} + \sum\limits_{i = 1}^3 {{{({h_i}d{x^i} - \alpha {\beta^i}dt)}^2}} \, ,
\end{equation} 
where $\alpha= [ h_0^2 + \sum_{i = 1}^3 (h_i\omega _i)^2 ]^{1/2}$ is the lapse function and $\beta^i={h_i}{\omega _i}/{\alpha }$ is the shift vector, while $h_0^2=-g_{00}$, $h_i^2=g_{ii}$ and $h_i^2 \omega_i = -g_{i0} = -g_{0i}$ are the non-zero components of the metric. Notice that the shift vector can also be written as $\beta^i=\pm \sqrt{\alpha^2-h_0^2}/{\alpha }$, depending on the direction of the black hole rotation.
 Using the $3+1$ formalism, we can isolate the effects of the spacetime rotation, which is useful for studying plasmas around rotating compact objects. 
It is also particularly convenient to introduce a locally nonrotating frame, the so-called ``zero-angular-momentum-observer'' (ZAMO) frame \cite{Bardeen_1972}, where the line element is $d{s^2} =  - d{{\hat t}^2} + \sum\nolimits_{i = 1}^3 {{{(d{{\hat x}^i})}^2}}  = {\eta _{\mu \nu }}d{{\hat x}^\mu }d{{\hat x}^\nu }$, with $d\hat t = \alpha dt$ and $d{{\hat x}^i} = {h_i}d{x^i} - \alpha {\beta^i}dt$ (here and in the following, quantities observed in the ZAMO frame are denoted with hats). For observers in the ZAMO frame, the spacetime is locally Minkowskian. Using this frame, any equation with the form of Eqs.~\eqref{eqmom} and \eqref{eqOhm} can be cast into the general form $\nabla_\nu S^{\mu\nu}=H^\mu$. For example, $S^{\mu\nu}=\mathfrak{h}U^\mu U^\nu$ for the energy-momentum equation and $S^{\mu\nu}=0$ for the Ohm's law, while the form of $H^{\mu}$ can be deduced from Eqs.~\eqref{eqmom} and \eqref{eqOhm}.
 This general equation can then be written in the ZAMO frame in vectorial form as \cite{Koide_2010}
\begin{eqnarray}\label{eqGen}
&&\frac{1}{h_1 h_2 h_3}\sum\limits_{j = 1}^3 \frac{\partial}{\partial x^j}\left[\frac{\alpha h_1h_2h_3}{h_j}\left(\hat S^{ij} +\beta^j \hat S^{i0} \right)\right]+\frac{\hat S^{00}}{h_i}\frac{\partial\alpha}{\partial x^i}\nonumber\\
&&-\alpha \sum\limits_{j = 1}^3 \left(G_{ij}\hat S^{ij}-G_{ji}\hat S^{jj}+\beta^jG_{ij}\hat S^{0i}-\beta^j G_{ji}\hat S^{0j} \right)\nonumber\\
&&+\sum\limits_{j = 1}^3\sigma_{ji}\hat S^{0j}=\alpha \hat H^i\, ,
\end{eqnarray}
with $i=1,2,3$, and where ${G_{ij}} =  - \left( {1/{h_i}{h_j}} \right)\partial {h_i} /\partial {x^j}$, and ${\sigma _{ij}} =  - \left( {1/{h_j}} \right)\partial\left( {\alpha {\beta ^i}} \right) /\partial {x^j}$. Vectors and tensors observed by the ZAMO frame are related to the covariant vectors and tensors as $\hat S^{00}=\alpha^2 S^{00}$, $\hat S^{0j}=\alpha h_j S^{0j}-\beta^j \hat S^{00}$, and $\hat S^{ij}=h_ih_j S^{ij}-\beta^i\hat S^{0j}-\beta^j\hat S^{0i}-\beta^i\beta^j\hat S^{00}$. 
Analogously, Maxwell equations can also been written in the ZAMO frame (see Ref.~\cite{Koide_2010}).

Since we are interested in analyzing magnetic reconnection around black holes, we consider the spacetime $\left( {{x^0},{x^1},{x^2},{x^3}} \right) = \left( {t,r,\theta ,\phi } \right)$ given by the Kerr metric \cite{Weinberg_72}, for which 
\begin{eqnarray}\label{}
h_0 = (1 - 2{r_g}r/{\Sigma})^{1/2}  \, , \quad h_1 = ({\Sigma}/{\Delta })^{1/2} \, , \nonumber \\
h_2 = \Sigma^{1/2} \, , \quad h_3 = (A/{\Sigma})^{1/2} \sin \theta \, , \; \; \quad \\
\omega_1 = \omega_2 = 0 \, , \quad \omega_3 = {{2r_g^2ar}}/{\Sigma} \, . \; \; \qquad  \nonumber
\end{eqnarray}
Here, $r_g = GM$ is the gravitational radius ($G$ is the gravitational constant and $M$ is the mass of the compact object) and $a=J/J_{\max}\leq 1$ is the rotation parameter ($J$ is the angular momentum and $J_{\max}=GM^2$). Moreover, $\Sigma  = {r^2} + {\left( {a{r_g}} \right)^2}{\cos ^2}\theta $, $\Delta  = {r^2} - 2{r_g}r + {\left( {a{r_g}} \right)^2}$, and $A = \big[ {{r^2} + {{\left( {a{r_g}} \right)}^2}} \big]^2 - \Delta {\left( {a{r_g}} \right)^2}{\sin ^2}\theta $. Finally, $\alpha  = (\Delta \Sigma /A)^{1/2}$ and $\beta^j=\beta^\phi \delta^{j\phi}$, where $\beta^\phi=h_3\omega_3/\alpha$ is a measurement of the rotation of this spacetime. Note that in Kerr spacetime we have $\beta^jG_{ij}\equiv 0$.

In our analysis, we adopt a Sweet-Parker-like approach \cite{Kulsrud_2005}, i.e., we look at magnetic 
reconnection under quasi-stationary conditions ($\partial_t \approx 0$) within narrow ($\delta \ll L$) quasi-two-dimensional current sheets. Quasi-stationarity is generally satisfied not only in steady-state, but also at the time of maximum reconnection rate. Current sheets where magnetic diffusion takes place can form in different locations around black holes. Here, we consider two paradigmatic cases with current sheets in the equatorial plane ($\theta=\pi/2$) of the massive body. We assume that these current sheets are in a stable orbit around the Kerr black hole \cite{abramo}.

\begin{figure}[]
\begin{center}
\includegraphics[bb = 0 0 291 254, width=8.5cm]{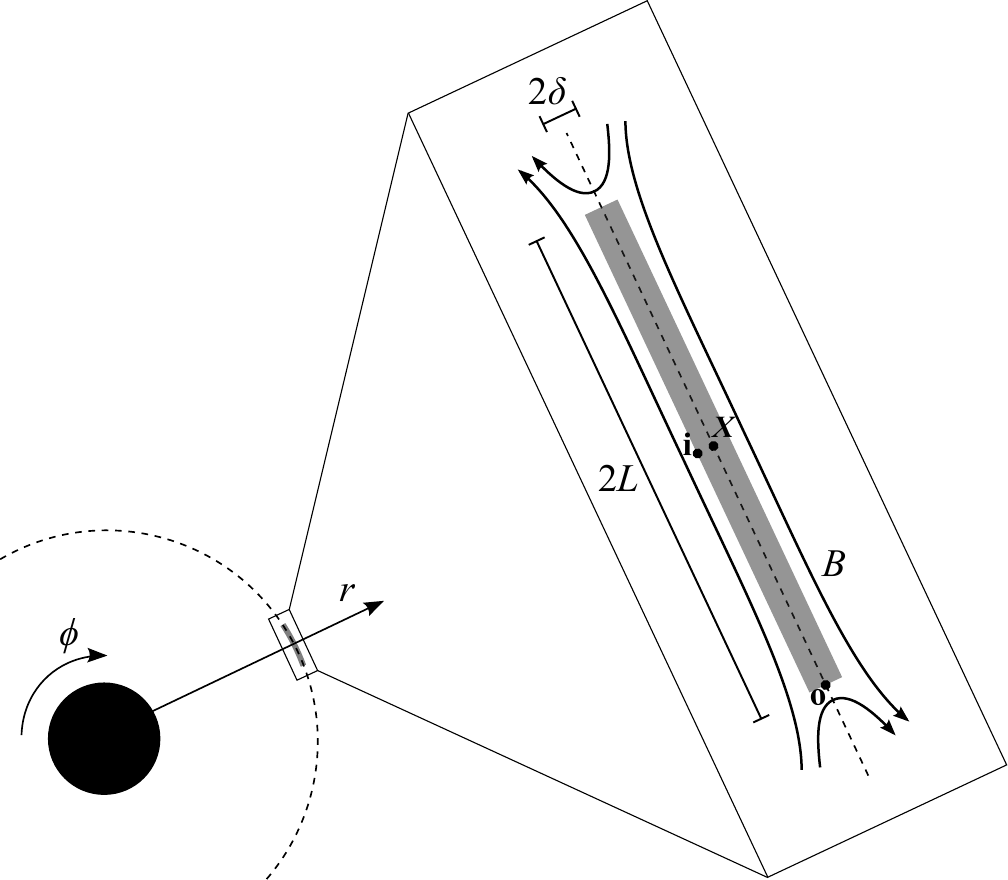}
\end{center}
\caption{Sketch of a reconnection layer in the azimuthal direction. The rotating black hole is represented by the black circle, while the magnetic diffusion region is marked by the shaded area.}
\label{fig1}
\end{figure}

{\it Reconnection layer in azimuthal direction.-}  We first consider a current sheet in the azimuthal direction, as shown in Fig.~\ref{fig1}. The magnetic field just upstream of this sheet is in the $\phi$-direction and has magnitude $\hat B_0$. In the diffusion region, $\hat v^r$ vanishes at the neutral line (where $\hat B^\phi = 0$). 
Similarly, at the neutral line $\hat J^\phi\approx 0\approx\hat J^r$, implying that $\rho_e'$ vanishes. We assume $\hat v^\theta \approx 0$, $\hat B^\theta \approx 0$, and that spatial variations of the fields with respect to $\theta$ (latitude) are negligible, i.e., $\partial_\theta \approx 0$. 
The same configuration has been adopted by Koide and Arai \cite{Koide_2010K} to examine the possibility of energy extraction from a rotating black hole. As in their work, we assume that the plasma in this configuration rotates in a circular orbit with Keplerian velocity.

We can calculate the outflow velocity from the diffusion region, $\hat v_{\rm{o}}$, from the spatial $\phi$-component of the energy-momentum equation \eqref{eqmom}. This equation can be written in the ZAMO frame by using Eq.~\eqref{eqGen}. Then, evaluating along the neutral line we get
\begin{equation}\label{mom3}
\frac{1}{\alpha h_1h_2}\frac{\partial}{\partial \phi} \left[\alpha h_1h_2 \mathfrak{h}\hat \gamma^2 \hat v^\phi\left(\hat v^\phi+\beta^\phi\right)\right] =  -h_3 {\hat J}^\theta {\hat B}_r - \frac{\partial p}{\partial \phi}   \, ,
\end{equation}
where $\hat \gamma  = {(1 - {\hat v^2})^{ - 1/2}}$ is the Lorentz factor. 
The outflow velocity $\hat v_{\rm{o}}$ can be found by integrating Eq. \eqref{mom3} from $\phi_X$ ($X$-point) to $\phi_{\rm{o}}$ (outflow-point). For this calculation, we assume that the rotation of the black hole is slow compared to the outflow velocity, $\hat v_{\rm{o}} \gg \beta^\phi$. This assumption is justified \emph{a posteriori}.  Note that pressure balance across the layer gives $p_X \approx {\hat B_0^2}/2$, while magnetic flux conservation yields
\begin{equation} \label{flux_cons}
{{\hat B}_r}|_{\rm{o}} \approx \left. {\left( {\frac{{r{h_1}}}{{{h_3}}}} \right)} \right|_{\rm{o}} \frac{\delta }{L} {{\hat B}_0} \, ,
\end{equation} 
where the symbol $|_{\rm{o}}$ indicates that the relevant quantities are evaluated at the outflow-point. Strictly speaking,  the magnetic flux in the diffusion region is not fully conserved, but Eq.~\eqref{flux_cons} is an estimated approximation for the reconnected magnetic field. In principle, one can derive the Sweet-Parker reconnection rate without using this condition. Furthermore, from Maxwell's equations we can evaluate 
\begin{equation}\label{flux_cons22}
{{\hat J}^\theta }{|_{\rm{o}}} \approx  - \left.\frac{1}{h_1} \right|_{\rm{o}}\frac{{{{\hat B}_0}}}{\delta } \, .
\end{equation} 
Thereby, 
by using the above assumptions  and Eqs.~\eqref{flux_cons} and \eqref{flux_cons22}, we can finally integrate Eq.~\eqref{mom3} to obtain 
\begin{eqnarray} \label{eq_velout}
\hat \gamma_{\rm{o}}\hat v_{\rm{o}} \approx \frac{1}{2}\, .
\end{eqnarray}
Here, we have considered a relativistically hot plasma, i.e. $\mathfrak{h} \approx 4p$ \cite{Chandra1938}. From Eq. (\ref{eq_velout}) we have that the Lorentz factor of the plasma outflow is $\hat \gamma_{\rm{o}}\approx {\cal O}(1)$, 
which implies a mildly relativistic outflow velocity ${\hat v_{\rm{o}}} \approx {\cal O}(1)$, as in the flat spacetime limit \cite{lyu,luca1}. Note that $\hat v_{\rm{o}} \gg \beta^\phi\approx {2 a r_g^2}/{r^2}$ when the current sheet is far from the black hole.

We proceed further by seeking the solution for the  inflow velocity $\hat v_{\rm{i}}$, which is a measure of the reconnection rate, and the current sheet width $\delta$, which, by means of Eq. (\ref{flux_cons}), also leads to the reconnected magnetic field ${{\hat B}_r}|_{\rm{o}}$. For this purpose, it is useful to consider separately the inner region, where magnetic diffusion occurs, from the outer region, where the plasma moves with a transport velocity that preserves the magnetic connections between plasma elements \cite{Newcomb,pegoraroEPJ,asenjoComissoCon,asenjo2015}. In the flat spacetime limit, $\partial_t \approx 0$ and $\partial_\phi \approx 0$ imply that ${{\hat E}_\theta }$ is uniform and can be used as a matching condition for these two regions. In a more general case, ${{\hat E}_\theta }$ can change in the equatorial plane, but for our purpose we only need to consider ${{\hat E}_\theta }$ at the $X$-point and the inflow point. From Maxwell's equations we have ${\partial_r} (\alpha {h_2}{{\hat E}_\theta }) = 0$ along the inflow line passing through the $X$-point, implying that ${\hat E}_\theta |_{\rm{i}} \approx {\hat E}_\theta |_X$ if the current sheet width $\delta$ is small, as can be seen \emph{a posteriori}.

Inside the current sheet, the $\theta$-component of Eq.~\eqref{eqOhm} written in the ZAMO frame reduces to
\begin{eqnarray}\label{eqOhm2}
{ {{\hat \gamma} {{\hat E}_\theta } + {\hat \gamma}{{\hat v}^\phi }{{\hat B}_r} } } = \eta {{\hat J}^\theta }  \, .
\end{eqnarray}
When evaluated at the $X$-point, this equation simply is ${\hat E}_\theta |_X = \eta \hat J^\theta$. 
On the other hand, outside the current sheet the plasma response to the electromagnetic field is well described by the ideal Ohm's law. Therefore, the rhs of Eq.~\eqref{eqOhm} can be neglected in this outer region. This implies that the $\theta$-component of the electric field at the inflow-point becomes just ${\hat E}_\theta |_{\rm{i}} \approx \hat v_{\rm{i}} \hat B_0$.

The final step requires the estimation of $\delta$,
which can be achieved from flux conservation. The inflow flux $\partial_r(\alpha h_2 h_3\hat\gamma_i \hat\rho \hat v_{i})/(h_1h_2h_3) \approx \alpha \hat\rho \hat\gamma_i \hat v_i /(h_1\delta)$
must balance the outflow flux $\partial_\phi(\alpha h_1 h_2\hat\gamma_o \hat\rho \hat v_{o})/(h_1h_2h_3) \approx \alpha  r \hat\rho \hat\gamma_o \hat v_o /(h_3 L)$. This leads us to find
\begin{equation}
\delta  \approx {\left. {\left( {\frac{{{h_3}}}{{{h_1}r}}} \right)} \right|_{\rm{o}}}\,\frac{{\hat \gamma}_{\rm{i}}\hat v_{\rm{i}}}{{\hat \gamma}_{\rm{o}}\hat v_{\rm{o}}}L  \, ,
\end{equation}
and, finally, the inflow velocity (reconnection rate)
\begin{equation}\label{RRcase1}
{\hat v_{\rm{i}}}\approx S^{-1/2}\left( \left.\frac{r}{h_3}\right|_{\rm{o}} \right)^{1/2} \, .
\end{equation}
Here, $S \equiv L/\eta$ is the relativistic Lundquist number ($c=1$ in our units), which compares the dynamical and resistive diffusion timescales of the plasma. We have considered ${\hat \gamma}_{\rm{i}}$ of the order of unity as $S\gg1$ for astrophysical plasmas.
The spacetime curvature leads to ${h_3}/{r}|_{\rm{o}}>1$, implying a lower reconnection rate with respect to the flat spacetime limit. This can be seen more explicitly in the scenario where the diffusion region is sufficiently far from the black hole, where the reconnection rate can be approximated as
\begin{equation}\label{rratepolar}
{\hat v_{\rm{i}}}\approx {S^{-1/2}}\left(1-\frac{a^2r_g^2}{4r_{\rm{o}}^2}\right) \, ,
\end{equation}
with $r_{\rm{o}}=\left.r\right|_{\rm{o}}$. We clearly see that the spacetime curvature effects lead to a decrease of the reconnection rate in this configuration. The responsible factor is the rotation of the black hole, while the curvature created by the black hole mass itself does not play an important role if the current sheet is in the azimuthal direction far from the black hole.

\begin{figure}[]
\begin{center}
\includegraphics[bb = 0 0 289 194, width=8.5cm]{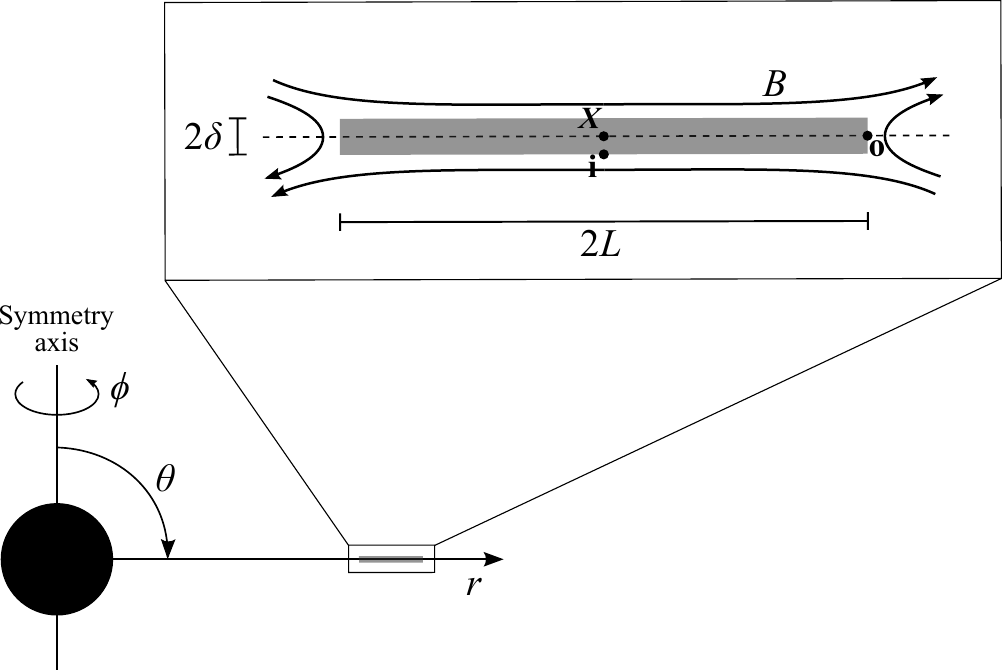}
\end{center}
\caption{Sketch of a reconnection layer in the radial direction. The rotating black hole is represented by the black circle, while the magnetic diffusion region is marked by the shaded area.}
\label{fig2}
\end{figure}

{\it  Reconnection layer in radial direction.-} We now consider a  magnetic reconnection configuration in which a narrow current sheet is located in the radial direction, as shown in Fig.~\ref{fig2}. The magnetic field just upstream of the current sheet is in the $r$-direction and has magnitude $\hat B_0$. In the diffusion region, the $\theta$-component of the velocity vanishes at the neutral line (where $\hat B^r = 0$). We assume $\hat v^\phi \approx 0 \approx \hat B^\phi$ and $\partial_\phi \approx 0$. This implies that $\rho'_e$ vanishes.
This configuration is particularly relevant for the split-monopole magnetic field geometry, where reconnection layers form in the radial direction \cite{LyutMcK_2011}.

In this case, the outflow velocity $\hat v_{\rm{o}}$ can be calculated by considering the spatial $r$-component of the energy-momentum equation \eqref{eqmom}. Writing it in the ZAMO frame \eqref{eqGen}, and evaluating it along the neutral line, we get
\begin{eqnarray}\label{mom2}
&&\frac{\partial}{\partial r}\left[\alpha h_2h_3 \mathfrak{h}\hat \gamma^2(\hat v^r)^2\right]+h_2h_3\frac{\partial\alpha}{\partial r}\mathfrak{h}\gamma^2=\nonumber\\
&&\qquad\qquad\qquad -\alpha {h_2}{h_3}\left( {\frac{{\partial p}}{{\partial r}} + h_1 {{\hat J}^\phi }{{\hat B}_\theta }} \right) \, .
\end{eqnarray}
In the integration of this equation from $r_X$ to $r_{\rm{o}}$ (in order to obtain the outflow velocity), we assume that the gravitational tidal field is negligible. 
This implies that the gravitational effects on different zones of that region, i.e. at $r_X$ and $r_{\rm{o}}$, are essentially the same. This also allows us to assume that the current sheet does not fall toward the black hole, and it can be considered in a rotational equilibrium. The above assumptions are no longer valid if the length of the current sheet \cite{Note1} is comparable with the radial distance to the black hole or if the current sheet is too close to it. In that case, a more detailed approach must be used, which is left for future works. Here, we restrict ourselves to this simplified model in order to obtain the first approximated contribution of gravity to magnetic reconnection.
Thus, at the integration level, the contribution of the spacetime curvature can be evaluated at the outflow-point. The $\phi$-component of the current density and the $\theta$-component of the magnetic field at the outflow-point can be estimated from Maxwell's equations and magnetic flux conservation. From them we obtain
\begin{equation}\label{radialJcong1}
{{\hat J}^\phi }{|_{\rm{o}}} \approx  - {\left. {\left[ {\frac{{{\partial _\theta }(\alpha {h_1}{{\hat B}_r})}}{{\alpha {h_1}{h_2}}}} \right]} \right|_{\rm{o}}} \approx  - \frac{{{{\hat B}_0}}}{\delta }  \, 
\end{equation}
and
\begin{equation}\label{radialJcong2}
{{\hat B}_\theta }{|_{\rm{o}}} \approx {\left. {{\frac{1}{{{h_1}}}}} \right|_{\rm{o}}}\frac{\delta }{L}{{\hat B}_0} \, ,
\end{equation} 
where ${\hat B}_0$ is in the radial direction.
Then, considering  a relativistically hot plasma and pressure balance across the layer, $\hat B_0^2/2 \approx p_X$, from the integration of Eq.~\eqref{mom2} along $r$ we find (at the same order)
\begin{eqnarray}\label{veloOUT}
{\hat \gamma _{\rm{o}}}^2{\hat v_{\rm{o}}}^2+L\left.\frac{\partial\ln\alpha}{\partial r}\right|_{\rm{o}}  {\hat \gamma _{\rm{o}}}^2\approx 1 \, ,
\end{eqnarray}
where Eqs.~\eqref{radialJcong1} and \eqref{radialJcong2} have been used.
Hence, the plasma outflow is characterized by a Lorentz factor and  an outflow velocity given by
\begin{equation}\label{veloOUT2}
{\hat \gamma _{\rm{o}}}\approx\left(1+L\left.\frac{\partial\ln\alpha}{\partial r}\right|_{\rm{o}}  \right)^{-1/2} ,\quad
{\hat v_{\rm{o}}} \approx \left(\frac{1}{2}-\frac{L}{2}\left.\frac{\partial\ln\alpha}{\partial r}\right|_{\rm{o}}  \right)^{1/2}\, .
\end{equation}
In the flat spacetime limit ($\alpha\rightarrow 1$), we recover again a mildly relativistic outflow velocity \cite{lyu,luca1}.

From Maxwell's equations \cite{Koide_2010} we have ${\partial_\theta} (\alpha {h_3}{{\hat E}_\phi}) = 0$ along the inflow line passing through the $X$-point, implying that ${\hat E}_\phi |_{\rm{i}} \approx {\hat E}_\phi |_X$ for a small $\delta$. Therefore, we can match the electric field considering the Ohm's law inside and outside the current sheet at the inflow point. Inside the current sheet, the $\phi$-component of Eq.~\eqref{eqOhm} written in the ZAMO frame formalism \eqref{eqGen} becomes
\begin{eqnarray}\label{}
{ {{\hat \gamma} {{\hat E}_\phi } + {\hat \gamma} {{\hat v}^r}{{\hat B}_\theta }}} = \eta {{\hat J}^\phi }  \, .
\end{eqnarray} 
At the $X$-point, this equation gives ${\hat E}_\phi |_{X} = {\eta} {\hat J}^\phi$. The $\phi$-component of the electric field of this equation has to be matched with that coming from the ideal Ohm's law evaluated at the inflow-point, i.e., ${{\hat E}_\phi }|_{\rm{i}} \approx {{\hat v}_{\rm{i}}}{{\hat B}_0}$. This leads us to find the inflow velocity once that the current sheet width is evaluated. From the balance between the energy inflow and outflow, we obtain 
\begin{equation}
\delta  \approx {\left. {{h_1}} \right|_{\rm{o}}}\frac{{\hat \gamma}_{\rm{i}}{\hat v}_{\rm{i}}}{{\hat \gamma}_{\rm{o}}{\hat v}_{\rm{o}}}L  \, .
\end{equation}
Therefore, the inflow velocity (reconnection rate) turns out to be
\begin{equation}\label{RRcase1}
\hat v_{\rm{i}} \approx  \frac{1}{\sqrt {S h_1|_{\rm{o}}}}     \left[\frac{ 1- L\left.\partial_r \ln \alpha \right|_{\rm{o}}}{1+ L\left.\partial_r\ln \alpha \right|_{\rm{o}}} \right]^{1/4} \, ,
\end{equation}
where again ${\hat \gamma}_{\rm{i}}$ is of unity order because $S\gg 1$.
Eq.~\eqref{RRcase1} contains the contribution of the gravitational field, from where we can clearly see that the spacetime curvature of the Kerr black hole modifies the reconnection rate. If we consider the ordering $r_g \ll r_{\rm{o}}$, the reconnection rate from Eq.~\eqref{RRcase1} can be approximated as
\begin{equation}\label{rrateazimu}
{\hat v_{\rm{i}}} \approx S^{-1/2}\left[1-\frac{r_g}{2r_{\rm{o}}} +\frac{(2 a^2-3)r_g^2}{8r_{\rm{o}}^2}\right]\, .
\end{equation}
This expression shows that the curvature created by the black hole mass leads to a decrease of the reconnection rate. On the contrary, the black hole rotation produces the opposite effect of increasing the reconnection rate. We observe, however, that these effects are small in the regime of validity of Eq.~\eqref{rrateazimu}.

{\it Conclusions.-} The presented analysis has allowed us to calculate the reconnection rate and other important properties of the reconnection layer for two configurations in the equatorial plane of a rotating black hole. We have shown how the spacetime curvature modifies the magnetic reconnection process in comparison to the flat spacetime limit. The rotation of the black hole leads to a decrease of the reconnection rate if the reconnection layer is in the azimuthal configuration. On the other hand, it is the spacetime curvature due to the black hole mass that acts to decrease the reconnection rate if the current sheet is in the radial direction.

These results have been obtained for a plasma not close to the event horizon of the black hole, and in a stable orbit \cite{abramo}.  We have also assumed a small rotation rate of the black hole. In future studies these assumptions may be relaxed. 
We also observe that collisionless effects are important in plasmas around black holes, and thereby, it is expected that they can also couple to gravity and affect the reconnection rate. 
A poloidal configuration could also be examined, and the black hole rotation is expected to modify the reconnection rate in a similar fashion to Eq.~\eqref{rratepolar}. This is relevant for exploring reconnection events in the corona and jet launching regions of the black hole \cite{Koide2006,Gouveia,Dexter,Kadowaki,Singh,Khiali}.

We finally observe that the potential formation of extremely elongated current sheets would result in slow reconnection rates. However, this situation is prevented by the occurrence of the plasmoid instability \cite{Comisso_2016}, which leads to the breakup of the current sheet in a fractal-like fashion \cite{Shibata2001} that ends when the smallest elementary current sheets are sufficiently thick to avoid the plasmoid instability. Our results can then be applied to these elementary current sheets, which are the actual locations where the magnetic energy is converted into plasma particle energy. Furthermore, our analysis can also be useful to better understand fast reconnection driven by turbulence \cite{Kowal2009,kowalparticle2,Takamotoin}, where the Sweet-Parker
configuration occurs in several current sheets simultaneously due to the wandering of the magnetic field lines.

\begin{acknowledgments}
It is a pleasure to acknowledge fruitful discussions with Gabriele Brambilla, Luis Lehner, Russell Kulsrud, and Alexander Tchekhovskoy. FAA thanks Fondecyt-Chile  Grant No. 11140025.
L.C. is grateful for the hospitality of the Universidad Adolfo Ib\'a\~nez, where part of this work was done. 
\end{acknowledgments}

\end{document}